\documentclass[prl,twocolumn]{revtex4-1}
\usepackage{amsmath,amssymb,bm,graphicx,hyperref}
\allowdisplaybreaks[1]

\renewcommand\d{\partial}
\newcommand\grad{\bm{\nabla}}
\newcommand\+{\dagger}
\newcommand\<{\langle}
\renewcommand\>{\rangle}
\renewcommand\k{{\bm{k}}}
\newcommand\p{{\bm{p}}}
\newcommand\q{{\bm{q}}}
\renewcommand\r{{\bm{r}}}
\renewcommand\L{\mathcal{L}}
\newcommand\N{\mathbb{N}}

\begin{document}

\title{Semisuper Efimov effect of two-dimensional bosons at a three-body resonance}

\author{Yusuke Nishida}
\affiliation{Department of Physics, Tokyo Institute of Technology,
Ookayama, Meguro, Tokyo 152-8551, Japan}

\date{February 2017}

\begin{abstract}
Wave-particle duality in quantum mechanics allows for a halo bound state whose spatial extension far exceeds a range of the interaction potential.
What is even more striking is that such quantum halos can be arbitrarily large on special occasions.
The two examples known so far are the Efimov effect and the super Efimov effect, which predict that spatial extensions of higher excited states grow exponentially and double exponentially, respectively.
Here, we establish yet another new class of arbitrarily large quantum halos formed by spinless bosons with short-range interactions in two dimensions.
When the two-body interaction is absent but the three-body interaction is resonant, four bosons exhibit an infinite tower of bound states whose spatial extensions scale as $R_n\sim e^{(\pi n)^2/27}$ for a large $n$.
The emergent scaling law is universal and is termed a semisuper Efimov effect, which together with the Efimov and super Efimov effects constitutes a trio of few-body universality classes allowing for arbitrarily large quantum halos.
\end{abstract}

\maketitle

\section{Introduction}
In classical mechanics, no force acts on particles if they reside outside a range of the interaction potential.
However, quantum mechanics allows them to form a bound state even though their mean separation far exceeds the potential range.
This is a consequence of the celebrated wave-particle duality, which states that particles behave as spatially extended waves of probability amplitude~\cite{Born:1955}.
Such classically forbidden bound states are generally referred to as quantum halos, which have attracted broad and long-standing interest in physics~\cite{Jensen:2004,Kornilov:2015}.

\begin{table}[b]
\caption{\label{tab:universality_class}
Trio of few-body universality classes allowing for arbitrarily large quantum halos.
A scaling exponent $\gamma$ for each class is independent of potential details but depends on other system information.
See Table~\ref{tab:scaling_law} for results in prototypical systems.
Also shown is a hyperspherical potential leading to each scaling law.}
\begin{ruledtabular}
\begin{tabular}{rll}
Universality class & Scaling law & Hyperspherical pot. \\[2pt]\hline
Efimov & $R_n\sim e^{\pi n/\gamma}$ & $V(r)\sim\#/r^2$ \\
Semisuper Efimov & $R_n\sim e^{(\pi n/\gamma)^2}$ & $V(r)\sim\#/r^2\ln r$ \\
Super Efimov & $R_n\sim e^{e^{\pi n/\gamma}}$ & $V(r)\sim\#/r^2\ln^2r$
\end{tabular}
\end{ruledtabular}
\end{table}

What is even more striking is that quantum halos can be arbitrarily large on special occasions.
One such example is the Efimov effect, known since its discovery back in 1970~\cite{Efimov:1970}.
When two particles interact resonantly in the $s$-wave channel, three particles exhibit an infinite tower of bound states whose spatial extensions grow exponentially for higher excited states (see Table~\ref{tab:universality_class}).
While spinless bosons in three dimensions are the prototypical system for the Efimov effect, it has been revealed that the same class of universal scaling law emerges for many other systems, such as mass-imbalanced fermions in three dimensions~\cite{Efimov:1973,Castin:2010,Bazak:2017}, anyons in two dimensions~\cite{Nishida:2008a}, bosons in one dimension~\cite{Nishida:2010}, and in various mixed-dimensional systems~\cite{Nishida:2008b,Nishida:2011}.

Now, one may ask whether the Efimov effect is the only possible route leading to arbitrarily large quantum halos.
The answer turned out to be no only recently in 2013, when another class of universal scaling law termed the super Efimov effect was discovered~\cite{Nishida:2013}.
Here, when two particles in two dimensions interact resonantly in the $p$-wave channel, three particles exhibit an infinite tower of bound states whose spatial extensions grow double exponentially for higher excited states (see Table~\ref{tab:universality_class}).
This super Efimov effect was originally predicted for spinless fermions and was later extended to mass-imbalanced mixtures as well~\cite{Moroz:2014}.

\begin{table}[b]
\caption{\label{tab:scaling_law}
Universal scaling laws for spinless bosons and fermions in all dimensionality.
An interesting interplay among the quantum statistics, dimensionality, required resonant interaction, and emergent scaling law can be seen.}
\begin{ruledtabular}
\begin{tabular}{rrlc}
System & Interaction & Scaling law & Reference \\[2pt]\hline
Bosons in 1D & Four body & $R_n\sim e^{\pi n/1.247}$ & \cite{Nishida:2010} \\
Bosons in 2D & Three body & $R_n\sim e^{(\pi n)^2/27}$ & This Letter \\
Bosons in 3D & Two body & $R_n\sim e^{\pi n/1.006}$ & \cite{Efimov:1970} \\
Fermions in 1D & \multicolumn{3}{c}{$\cdots$} \\
Fermions in 2D & Two body & $R_n\sim e^{e^{3\pi n/4}}$ & \cite{Nishida:2013} \\
Fermions in 3D & \multicolumn{3}{c}{$\cdots$}
\end{tabular}
\end{ruledtabular}
\end{table}

In this Letter, we establish yet another new class of universal scaling law that allows for arbitrarily large quantum halos formed by spinless bosons with short-range interactions in two dimensions.
When the two-body interaction is absent but the three-body interaction is resonant in the $s$-wave channel~\cite{two-body}, four bosons exhibit an infinite tower of bound states whose spatial extensions scale as $R_n\sim e^{(\pi n)^2/27}$ for a large $n\in\N$.
The emergent scaling law is universal and qualitatively distinct from the previously known scaling laws, as seen in Table~\ref{tab:scaling_law}.
This novel phenomenon shall be termed a semisuper Efimov effect because its growth rate lies between the Efimov effect and the super Efimov effect, which together constitute a trio of few-body universality classes allowing for arbitrarily large quantum halos.

\section{Model analysis}
We first demonstrate the semisuper Efimov effect by a model analysis and discuss its universality later.
Let us consider spinless bosons in two dimensions without a two-body interaction but with a three-body interaction.
For simplicity, we shall employ the following model Hamiltonian:
\begin{align}
H &= \int\!\frac{d\k}{(2\pi)^2}\frac{\k^2}{2}\phi_\k^\+\phi_\k
- v_0\int\!\frac{d\k d\p d\q d\p'd\q'}{(2\pi)^{10}}\chi(\p,\q)\chi(\p',\q') \notag\\
&\quad \times
\phi^\+_{\frac\k3+\p'}\phi^\+_{\frac\k3-\frac{\p'}2+\q'}\phi^\+_{\frac\k3-\frac{\p'}2-\q'}
\phi_{\frac\k3-\frac\p2-\q}\phi_{\frac\k3-\frac\p2+\q}\phi_{\frac\k3+\p}.
\end{align}
Here, we set $\hbar=1$ and $m=1$ for a mass of bosons, and the three-body potential is assumed to be separable with the form factor provided by $\chi(\p,\q)=e^{-(3\p^2/4+\q^2)/2\Lambda^2}$, where $v_0>0$ is an attractive coupling constant and $\Lambda$ is a momentum cutoff.
This separable potential is advantageous in that three-body and four-body problems can be solved analytically to gain valuable insights into universal low-energy physics.

\begin{figure}[b]
\includegraphics[width=\columnwidth,clip]{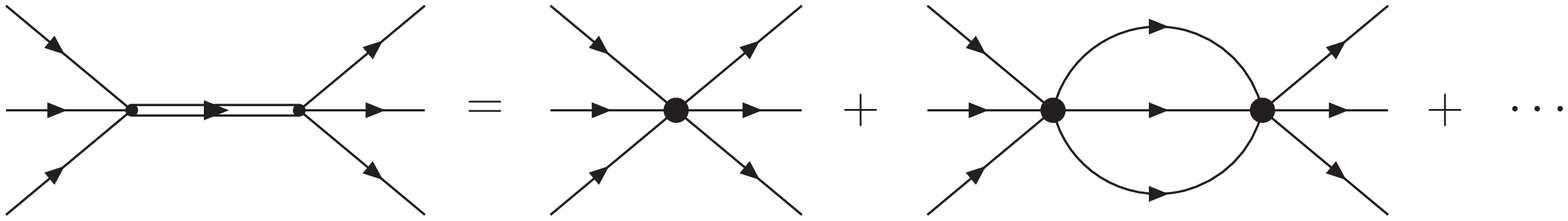}
\caption{\label{fig:3-body_amplitude}
Feynman diagrams for the three-body scattering $T$ matrix represented by the double line.}
\end{figure}

By summing an infinite series of Feynman diagrams depicted in Fig.~\ref{fig:3-body_amplitude}, the three-body scattering $T$ matrix from incoming $(\k/3+\p,\k/3-\p/2+\q,\k/3-\p/2-\q)$ to outgoing momenta $(\k/3+\p',\k/3-\p'/2+\q',\k/3-\p'/2-\q')$, with a total energy $\varepsilon$, is found to be
\begin{align}\label{eq:3-body_amplitude}
T_3(E,\k;\p,\q;\p',\q') = \frac{72\pi^2\chi(\p,\q)\chi(\p',\q')}
{2\pi^2/v_0-\Lambda^2-E\,e^{-E/\Lambda^2}E_1(-E/\Lambda^2)},
\end{align}
where $E\equiv\varepsilon-\k^2/6+i0^+$ is the collision energy in the center-of-mass frame and $E_1(w)\equiv\int_w^\infty\!dt\,e^{-t}/t$ is the first-order exponential integral.
The three-body resonance is defined by the divergence of the $T$ matrix at zero collision energy, which corresponds to a bound state of three bosons with zero binding energy, and is achieved by tuning the coupling constant at $v_0=2\pi^2/\Lambda^2$.

\begin{figure}[t]
\includegraphics[width=\columnwidth,clip]{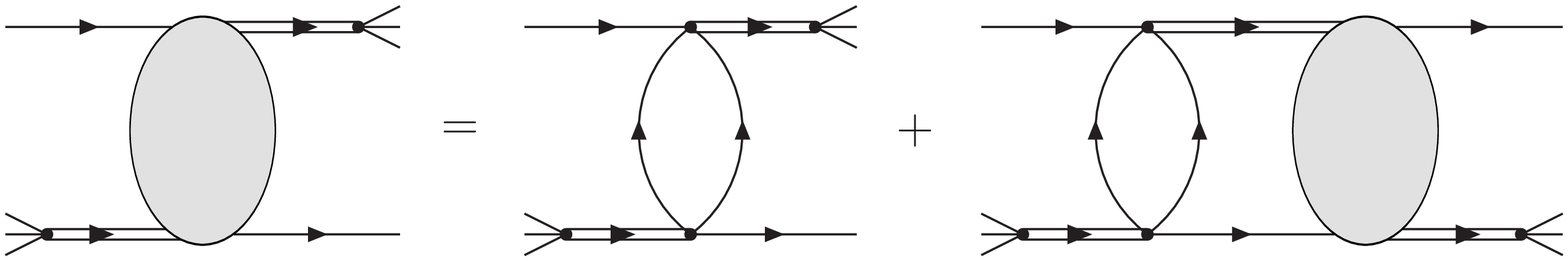}
\caption{\label{fig:4-body_amplitude}
Feynman diagrams for the four-body scattering $T$ matrix represented by the blob.}
\end{figure}

We now turn to a bound state problem of four bosons right at the three-body resonance.
The four-body scattering $T$ matrix denoted by $T_4(E;\p,\p')$ satisfies a Skornyakov--Ter-Martirosyan--type integral equation depicted in Fig.~\ref{fig:4-body_amplitude}, where $E$ is a collision energy in the center-of-mass frame and $\p$ ($\p'$) is an incoming (outgoing) relative momentum of one boson with respect to the others.
When the collision energy approaches one of the four-body binding energies, $E\to-\kappa^2<0$, the $T$ matrix factorizes as $T_4(E;\p,\p')\to Z(\p)Z^*(\p')/(E+\kappa^2)$, so that the resulting residue function satisfies
\begin{align}\label{eq:4-body_residue}
& Z(\p) = 9\pi\int\!\frac{d\p'}{(2\pi)^2} \notag\\
& \times \frac{e^{[(\p^2+\p'^2)/3+\kappa^2]/\Lambda^2}
E_1[(\frac{3\p^2+3\p'^2+2\p\cdot\p'}{4}+\kappa^2)/\Lambda^2]}
{(\frac{2\p'^2}{3}+\kappa^2)\,e^{(2\p'^2/3+\kappa^2)/\Lambda^2}
E_1[(\frac{2\p'^2}{3}+\kappa^2)/\Lambda^2]}\,Z(\p').
\end{align}
Because different partial-wave channels are decoupled, we can set $Z(\p)=e^{i\ell\varphi_\p}z(|\p|)$ and focus on the $\ell=0$ channel, which is most favorable to bound states because of the absence of a centrifugal barrier.

The integral equation (\ref{eq:4-body_residue}) can be solved analytically in the low-energy limit $\kappa\ll\Lambda$ with the leading-logarithm approximation~\cite{Nishida:2013,Moroz:2014}.
To this end, we assume that the integral is dominated by the region $\kappa\ll p'\ll\Lambda$ and split the integral into two parts, $\kappa\ll p'\ll p$ and $p\ll p'\ll\Lambda$, where the sum of $p^2$ and $p'^2$ in the integrand is replaced with whichever is larger.
Consequently, the integral equation (\ref{eq:4-body_residue}) is reduced to
\begin{align}\label{eq:reduced_integral}
\zeta(x) = \frac{27}{4}\int_\delta^x\!dx'\zeta(x')
+ \frac{27}{4}\int_x^{\ln\Lambda/\kappa}\!dx'\frac{x}{x'}\zeta(x'),
\end{align}
where $x^{(\prime)}\equiv\ln\Lambda/p^{(\prime)}$ and $\zeta(x)\equiv z(p)$ are introduced and $\delta>0$ is a constant sensitive to ultraviolet physics.
Then, by differentiating both sides of Eq.~(\ref{eq:reduced_integral}) with respect to $x$ twice, it can be brought into a differential equation,
\begin{align}
\zeta''(x) = -\frac{27}{4}\frac{\zeta(x)}{x},
\end{align}
which is solved by $\zeta(x)=A\sqrt{x}\,J_1(\sqrt{27x})+B\sqrt{x}\,Y_1(\sqrt{27x})$.
The resulting solution must satisfy two boundary conditions obtained by setting $x=\delta$ and $\ln\Lambda/\kappa$ in Eq.~(\ref{eq:reduced_integral}):
\begin{subequations}
\begin{align}\label{eq:boundary1}
\zeta(\delta) &= \frac{27}{4}\int_\delta^{\ln\Lambda/\kappa}\!dx'\frac\delta{x'}\zeta(x'), \\\label{eq:boundary2}
\zeta(\ln\Lambda/\kappa) &= \frac{27}{4}\int_\delta^{\ln\Lambda/\kappa}\!dx'\zeta(x').
\end{align}
\end{subequations}
In the low-energy limit $\ln\Lambda/\kappa\to\infty$, the first boundary condition (\ref{eq:boundary1}) merely relates $B/A$ to $\delta$, while the second boundary condition (\ref{eq:boundary2}) leads to $\tan(\sqrt{27\ln\Lambda/\kappa})=(B-A)/(B+A)$.
Therefore, we find an infinite tower of bound state solutions whose binding energies $E_n=\kappa_n^2$ for a large $n\in\N$ are provided by
\begin{align}\label{eq:semisuper_Efimov}
E_n \propto e^{-2(\pi n+\theta)^2/27}\Lambda^2,
\end{align}
where $\theta\equiv\arctan[(B-A)/(B+A)]$ is a nonuniversal constant defined modulo $\pi$.
This is nothing short of the semisuper Efimov effect.

\begin{table}[t]
\caption{\label{tab:binding_energy}
Up to 80th four-body binding energies in the form of $\ln\Lambda/\kappa_n$, which gradually converge to the universal scaling law; $\lim_{n\to\infty}(\sqrt{\ln\Lambda/\kappa_n}-\sqrt{\ln\Lambda/\kappa_{n-1}})=\pi/\sqrt{27}\approx0.604600$.}
\begin{ruledtabular}
\begin{tabular}{crcrc}
& $n$ && $\ln\Lambda/\kappa_n$ & $\sqrt{\ln\Lambda/\kappa_n}-\sqrt{\ln\Lambda/\kappa_{n-1}}$ \\[2pt]\hline
& 0 && $-$0.27549 & $\cdots$ \\
& 1 && 2.12031 & $\cdots$ \\
& 2 && 5.02648 & 0.785853 \\
& 3 && 8.65786 & 0.700444 \\
& 4 && 13.0241 & 0.666470 \\
& 5 && 18.1231 & 0.648229 \\
& 10 && 54.5866 & 0.618347 \\
& 20 && 182.339 & 0.608565 \\
& 40 && 657.162 & 0.605676 \\
& 80 && 2484.10 & 0.604881
\end{tabular}
\end{ruledtabular}
\end{table}

Our finding above can also be confirmed by numerically solving the integral equation (\ref{eq:4-body_residue}) in the $\ell=0$ channel.
The obtained binding energies of four bosons are shown in Table~\ref{tab:binding_energy}, which indeed converge to the universal scaling law (\ref{eq:semisuper_Efimov}), with three digits of precision up to the 80th bound state.
Because the leading-logarithm approximation does not provide any nontrivial solutions for $\ell\neq0$, arbitrarily large quantum halos do not emerge in the other partial-wave channels.

\section{Universality}
With our model analysis completed, two questions naturally arise.
Is the semisuper Efimov effect universal?
What if a two-body interaction is present?
A valuable insight for answering these questions can be gained from Eq.~(\ref{eq:3-body_amplitude}).
Here, we find that the three-body scattering $T$ matrix right at the three-body resonance in the low-energy limit $E\to0$ has exactly the same form as a propagator of a particle whose mass is $3m$ up to overall logarithmic energy dependence.
This speciality is actually common to the resonant $s$-wave scattering in four dimensions~\cite{Nishida:2006,Nishida:2007} and the resonant $p$-wave scattering in two dimensions~\cite{Nishida:2008a,Nishida:2013,Moroz:2014}, which can be attributed to the logarithmic divergence of wave function normalization integrals at origin~\cite{Nussinov:2006}.
Consequently, the low-energy limit of the resonant three-body scattering in two dimensions is universally described by a propagation of a pointlike trimer.

It is then straightforward to arrive at the following effective field theory to capture low-energy physics of two-dimensional spinless bosons close to the three-body resonance:
\begin{align}
\L &= \phi^\+\biggl(i\d_t+\frac{\grad^2}2\biggr)\phi + v_2\phi^\+\phi^\+\phi\phi \notag\\
&\quad + \Phi^\+\biggl(i\d_t+\frac{\grad^2}6-\varepsilon_0\biggr)\Phi
+ g\Phi^\+\phi\phi\phi + g\phi^\+\phi^\+\phi^\+\Phi \notag\\[2pt]
&\quad + v_4\phi^\+\Phi^\+\Phi\phi + v_6\Phi^\+\Phi^\+\Phi\Phi.
\end{align}
This Lagrangian density is built so as to include all relevant and marginal couplings written in terms of the boson $\phi$ and trimer $\Phi$ fields because irrelevant couplings disappear in the low-energy limit.
The three-body resonance is achieved by tuning the bare detuning parameter at $\varepsilon_0=g^2\Lambda^2/2\pi^2$, while the other couplings are dimensionless and acquire logarithmic energy dependence through their renormalization.
Because our interest lies in the bound state problem of four bosons, we shall not consider the six-body coupling $v_6$ further.

\begin{figure}[b]
\includegraphics[width=0.9\columnwidth,clip]{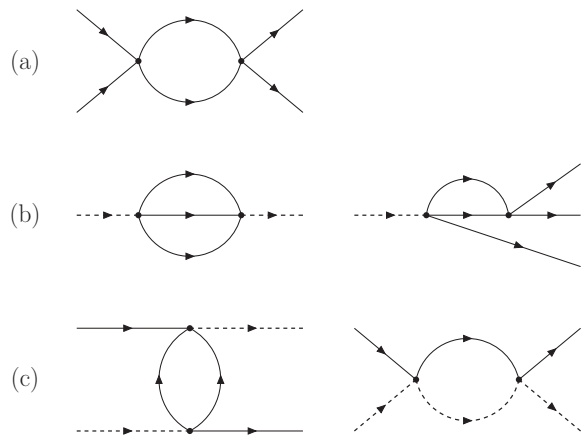}
\caption{\label{fig:renormalization}
Feynman diagrams to renormalize the (a) two-body, (b) three-body, and (c) four-body couplings.
The solid and dashed lines represent propagators of the boson and trimer fields, respectively.}
\end{figure}

The two-body $v_2$, three-body $g$, and four-body $v_4$ couplings are renormalized by Feynman diagrams depicted in Fig.~\ref{fig:renormalization}, so that their running at a momentum scale $\kappa\equiv e^{-s}\Lambda$ is governed by
\begin{subequations}
\begin{align}
\frac{dv_2}{ds} &= \frac{v_2^2}{\pi}, \\\label{eq:RG_3-body}
\frac{dg}{ds} &= -\frac{g^3}{2\pi^2} + \frac{3gv_2}{\pi}, \\\label{eq:RG_4-body}
\frac{dv_4}{ds} &= \frac{9g^2}{\pi} + \frac{3v_4^2}{4\pi} - \frac{g^2v_4}{\pi^2} + O(g^2v_2).
\end{align}
\end{subequations}
While the renormalization group (RG) equations for $v_2$ and $g$ are exact, that for $v_4$ suffers from higher-order corrections $\sim g^2v_2^i$, with $i\geq1$, but they do not interfere with our discussion below.
Low-energy physics of the system under consideration can be deduced by studying how these running couplings behave in the infrared limit $s\to\infty$.

When the two-body interaction is absent, i.e., $v_2=0$, corresponding to an infinite scattering length in two dimensions, the RG equation for $g$ is solved by
\begin{align}
g^2(s) = \frac1{\frac1{g^2(0)}+\frac{s}{\pi^2}}.
\end{align}
Consequently, the three-body coupling is marginally irrelevant, in that it decreases as $g^2\to\pi^2/s$ toward the infrared limit $s\to\infty$.
By substituting this asymptotic form of $g^2$ into Eq.~(\ref{eq:RG_4-body}), the solution to the RG equation for $v_4$ in the same limit is obtained as
\begin{align}\label{eq:4-body_coupling}
v_4(s) \to -\frac{6\pi}{\sqrt{3s}}\cot(\sqrt{27s}-\theta),
\end{align}
where $\theta$ is a nonuniversal constant depending on initial conditions for $g$ and $v_4$ at the ultraviolet scale $s\sim0$.
Therefore, we find that $\sqrt{s}v_4$ is a periodic function of $\sqrt{s}$ and diverges at $s_n=(\pi n+\theta)^2/27$ for a large $n\in\N$.
These divergences of the four-body coupling in its renormalization group flow indicate the existence of an infinite tower of characteristic energy scales, $E_n\sim e^{-2s_n}\Lambda^2$, in the system of four bosons.
The identification of these scales with four-body binding energies leads to our prediction of the semisuper Efimov effect presented in Eq.~(\ref{eq:semisuper_Efimov}).
Furthermore, our RG analysis demonstrates the universality of the semisuper Efimov effect because no assumption on microscopic details of the short-range two-body and three-body potentials has been made.

On the other hand, when the two-body interaction with a finite scattering length is present, the RG equation for $v_2$ is solved by
\begin{align}
v_2(s) = \frac1{\frac1{v_2(0)}-\frac{s}{\pi}},
\end{align}
whose magnitude decreases as $v_2\to-\pi/s$ toward the infrared limit $s\to\infty$.
By substituting this asymptotic form of $v_2$ into Eq.~(\ref{eq:RG_3-body}), its solution in the same limit is obtained as $g\to C/s^3$, where $C$ is a nonuniversal constant.
Because the three-body coupling rapidly vanishes, all terms proportional to $g^2$ in Eq.~(\ref{eq:RG_4-body}) can be ignored, so that the four-body coupling turns out to behave monotonically as $v_4\to-4\pi/3s$.
Therefore, we find that the semisuper Efimov effect as manifested by Eq.~(\ref{eq:4-body_coupling}) disappears in the presence of the two-body interaction.

\section{Concluding remarks}
We hereby established that arbitrarily large quantum halos can be formed by four spinless bosons in two dimensions when the two-body interaction is absent but the three-body interaction is resonant~\cite{two-body}.
Their spatial extensions scale as $R_n\sim1/\sqrt{mE_n}\propto e^{(\pi n+\theta)^2/27}/\Lambda$ for higher excited states, which is the new class of universal scaling law termed the semisuper Efimov effect.
Because it disappears in the presence of the two-body interaction, both two-body and three-body interactions need to be tuned simultaneously to achieve the semisuper Efimov effect.
Such double fine-tuning is generally challenging, but it may be possible in ultracold atom experiments.
In fact, there have been a number of proposed schemes to implement vanishing two-body but strong three-body interactions both in a free space~\cite{Petrov:2014a} and in an optical lattice~\cite{Daley:2014,Petrov:2014b,Paul:2016}.
Once the three-body resonance without the two-body interaction is achieved in a two-dimensional Bose system~\cite{lattice_model}, the emergent semisuper Efimov states of four bosons may be observed in ultracold atom experiments through resonantly enhanced atom losses by varying a detuning parameter~\cite{Kraemer:2006,Ferlaino:2009,Huang:2014}.

Besides such experimental feasibility, our finding in this Letter significantly advances our perspective on quantum few-body physics, in particular, by completing the universal scaling laws for spinless bosons in all dimensionality~\cite{Efimov:1970,Nishida:2010}.
Together with that for spinless fermions in two dimensions~\cite{Nishida:2013}, an interesting interplay among the quantum statistics, dimensionality, required resonant interaction, and emergent scaling law can now be seen in Table~\ref{tab:scaling_law}.
The universal scaling laws discovered so far are classified into either the Efimov effect, the semisuper Efimov effect, or the super Efimov effect, which together constitute a trio of few-body universality classes allowing for arbitrarily large quantum halos.

The last important question is whether there exist even more few-body universality classes allowing for arbitrarily large quantum halos.
Some insight may be gained by recalling hyperspherical potentials leading to the Efimov effect~\cite{Efimov:1970}, the semisuper Efimov effect~\cite{Efremov:2014,Efremov:2015,mass_imbalance}, and the super Efimov effect~\cite{Volosniev:2014,Gao:2015}, which are $V(r)\sim\#/r^2$, $\#/r^2\ln r$, and $\#/r^2\ln^2r$, respectively.
These potential forms can be deduced from the semiclassical quantization condition, $\pi n\sim\int^{R_n}\!dr\sqrt{-V(r)}$, in the zero binding energy limit~\cite{Landau-Lifshitz}.
The resulting hierarchical pattern seen in Table~\ref{tab:universality_class}, together with the Coulomb potential leading to the Rydberg law, inspires us to suggest that the hyperspherical potentials of
\begin{subequations}
\begin{flalign}
&& V(r) &\sim \#/[r^2(\ln r)^2(\ln\ln r)] && \\
\text{and} \notag\\
&& V(r) &\sim \#/[r^2(\ln r)^2(\ln\ln r)^2], &&
\end{flalign}
\end{subequations}
if realized, lead to the new scaling laws of
\begin{subequations}
\begin{flalign}
&& R_n &\sim \exp\{\exp[(\pi n/\gamma)^2]\} && \\
\text{and} \notag\\
&& R_n &\sim \exp\{\exp[\exp(\pi n/\gamma)]\}, &&
\end{flalign}
\end{subequations}
respectively, for which a semihyper Efimov effect and a hyper Efimov effect should be reserved.
Whether such semihyper and hyper Efimov effects actually emerge in quantum few-body systems with short-range interactions remains to be addressed in future work.

\acknowledgments
This work was supported by JSPS KAKENHI Grants No.~JP15K17727 and No.~JP15H05855.


\begin{thebibliography}{99}
\bibitem{two-body}
To be precise, what is required by the absence of the two-body interaction in this Letter is not the identically vanishing two-body potential, but instead only the infinite scattering length in two dimensions corresponding to a bound state of two bosons with zero binding energy.
\bibitem{lattice_model}
For reference, in the case of the following Bose-Hubbard model on a square lattice,
$$H = -\frac{t}{2}\sum_{\<\r,\r'\>}b_\r^\+b_{\r'}
- \frac{U}{3!}\sum_\r b_\r^\+b_\r^\+b_\r^\+b_\r b_\r b_\r,$$
the three-body resonance is achieved at $U/t\approx5.10217$.
\end{thebibliography}
\end{document}